\documentstyle[11pt,newpasp,twoside]{article}
\newcommand\RRc       {candidate RR Lyrae stars}
   
\def\edcomment#1{\iffalse\marginpar{\raggedright\sl#1\/}\else\relax\fi}
\reversemarginpar

\begin{document}
\title{Reaching to the Edge of the Milky Way Halo with SDSS}

\author{\v{Z}eljko Ivezi\'{c}$^1$, Robert Lupton$^1$, David Schlegel$^1$, 
Vernesa Smol\v{c}i\'{c}$^{1,2}$, David Johnston$^3$, Jim Gunn$^1$, Jill Knapp$^1$, 
Michael Strauss$^1$, Constance Rockosi$^4$ (SDSS Collaboration)}

\affil{$^1$ Princeton University Observatory, Princeton, NJ 08544}
\affil{$^2$ University of Zagreb, Dept. of Physics, Bijeni\v{c}ka 
            cesta 32, 10000 Zagreb, Croatia}
\affil{$^3$ University of Chicago, Astronomy \& Astrophysics Center, 
            5640 S. Ellis Ave., Chicago, IL 60637}
\affil{$^4$ University of Washington, Dept. of Astronomy,
            Box 351580, Seattle, WA 98195}

\begin{abstract}
We discuss a sample of over 3000 \RRc\ selected by various methods
using Sloan Digital Sky Survey data for about 1000 deg$^2$ of sky. 
These stars probe the halo structure out to $\sim$100 kpc from
the Galactic center. Their spatial and radial velocity distributions 
are very inhomogeneous, with the most prominent features tracing the 
Sgr dwarf tidal stream.  Outside the Sgr dwarf tidal stream, the spatial 
distribution in the 5--60 kpc range of Galactocentric radius $R$ is well 
described by an $R^{-3}$ power law.
\end{abstract}

\section{Introduction}

\subsection{RR Lyrae Stars as a Probe of Galactic Halo}
Studies of substructures, such as clumps and streams, in the Galactic
halo can help constrain the formation history of the Milky Way (Helmi,
White \& Springel 2003, and references therein). Hierarchical models of 
galaxy formation predict that these substructures should be ubiquitous in
the outer halo, where the dynamical timescales are sufficiently long
for them to remain spatially and dynamically coherent (e.g. Johnston 
et al.~1996). Among the best tracers to study the outer halo are RR Lyrae 
stars because

\begin{itemize}
\item
They are nearly standard candles ($\sigma_M \sim 0.15$ mag) and thus 
it is straightforward to determine their distance
\item
They are sufficiently bright to be detected at large distances
(5-100 kpc for $14 < r < 20.7$).
\item
They are sufficiently numerous for detailed tracing of the halo 
structure (more so than e.g. C and M giants)
\end{itemize}

\subsection{The Selection of Candidate RR Lyrae Stars Using SDSS Data}
The Sloan Digital Sky Survey (SDSS; York et al. 2000, Azebajian et al.~2003, 
and references therein) is revolutionizing studies of the Galactic halo by
providing homogeneous and deep ($r < 22.5$) photometry in five
passbands ($u$, $g$, $r$, $i$, and $z$, Fukugita et al.~1996, Gunn et al.~1998) 
accurate to 0.02 mag (Ivezi\'{c} et al.~2003a); ultimately, up to 10,000 deg$^2$ 
of sky in the Northern Galactic Cap will be observed. The survey sky coverage 
will result in photometric measurements for over 100 million stars and a similar
number of galaxies. Astrometric positions are accurate to better than
0.1 arcsec per coordinate (rms) for sources with $r<20.5^m$
(Pier et al.~2003), and the morphological information from the images
allows robust star-galaxy separation to $r \sim$ 21.5$^m$ (Lupton
et al.~2001).

There are several methods for selecting candidate RR Lyrae stars using 
SDSS. While RR Lyrae stars are best found by obtaining 
well-sampled light curves, Ivezi\'{c} et al. (2000) 
demonstrated that candidates can be efficiently and robustly found 
even with two-epoch data, using the accurate multi-band SDSS photometry.
The robustness of their sample and efficiency estimates were later 
confirmed by the QUEST survey (Vivas et al. 2001).
Furthermore, even when multiple SDSS imaging observations are 
unavailable, candidate RR Lyrae stars can be selected by comparing SDSS 
imaging and spectro-photometric magnitudes (Smol\v{c}i\'{c} et al. 2003), 
SDSS imaging and POSS magnitudes (Sesar et al. 2003), and even by color 
selection from single-epoch data (Ivezi\'{c} et al. 2003b). In this contribution
we discuss over 3,000 candidate variable stars selected by comparing
SDSS imaging data to another imaging epoch, or to spectro-photometric
measurements, which are the most robust methods. For a subset of 700 of these
stars SDSS also provides radial velocity measurements.

\section{The Selection and Analysis of the SDSS RR Lyrae Candidates}

\subsection{Variability-selected Candidate RR Lyrae Stars }
We used multiple SDSS imaging data and SDSS spectrophotometric
measurements to select 3,127 RR Lyrae candidates in $\sim$1,000 deg$^2$
of sky. The details of the selection are described in detail by Ivezi\'{c}
et al. (2000) and Smol\v{c}i\'{c} et al. (2003). The sample completeness 
is about 35\%, and its efficiency
(the estimated fraction of true RR Lyrae stars) is about 90\%. 

The overall distribution of the selected candidates on the sky is shown
in Figure~1. The most significant feature, found at (R.A.$\sim$218$^o$,
Dec$\sim$0$^o$), is the well-known clump associated with the
Sgr dwarf tidal stream (Ivezi\'{c} et al. 2000, Yanny et al. 2000, 
Vivas et al. 2001). It is easily visible in Figure~2 which shows two
projections of the Galactic distribution for the same sample.
The clump is at (X$\sim$20 kpc, Y$\sim$-10 kpc, Z$\sim$40 kpc).

\begin{figure}
\plotfiddle{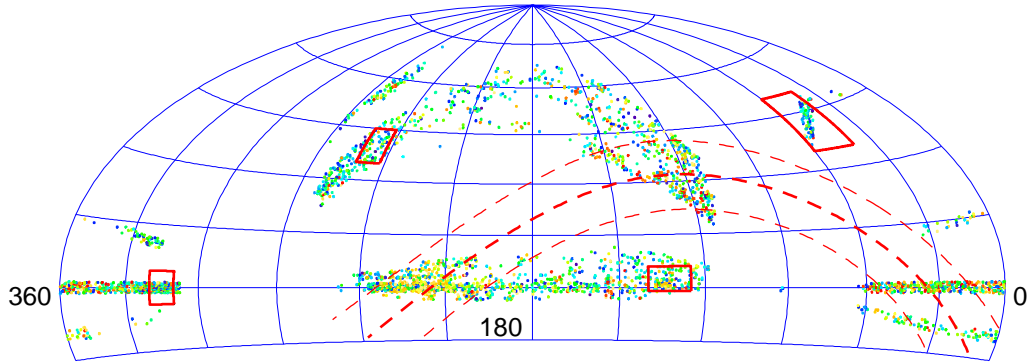}{5cm}{0}{74}{74}{-230}{-380}
\caption{The sky distribution in equatorial Aitoff projection of SDSS 
candidate RR Lyrae stars selected by variability (dots). 
The thick dashed line shows a great circle that corresponds to the Sgr 
dwarf tidal stream, as outlined by the distributions of C stars (Ibata 
et al. 2001) and M giants (Majewski et al. 2003). Thin dashed lines define
a 20 degree wide strip centered on this great circle. The polygons
define four pencil beams utilized for studying the distribution of 
stars outside the Sgr dwarf tidal stream.}
\end{figure}

\begin{figure}
\plotfiddle{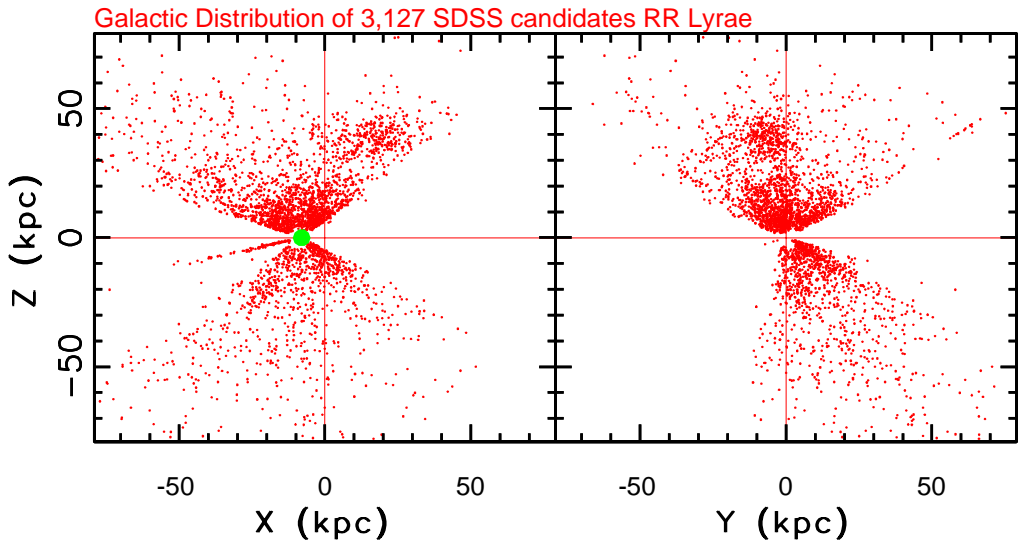}{7cm}{0}{74}{74}{-220}{-350}
\caption{The distribution of SDSS candidate RR Lyrae stars
selected by variability in the Galactocentric coordinate system.
The prominent clump at (X$\sim$20 kpc, Y$\sim$-10 kpc, Z$\sim$40 kpc)
belongs to the Sgr dwarf tidal stream.}
\end{figure}

\subsection{The Properties of the Sgr Dwarf Tidal Stream }

\begin{figure}
\plotfiddle{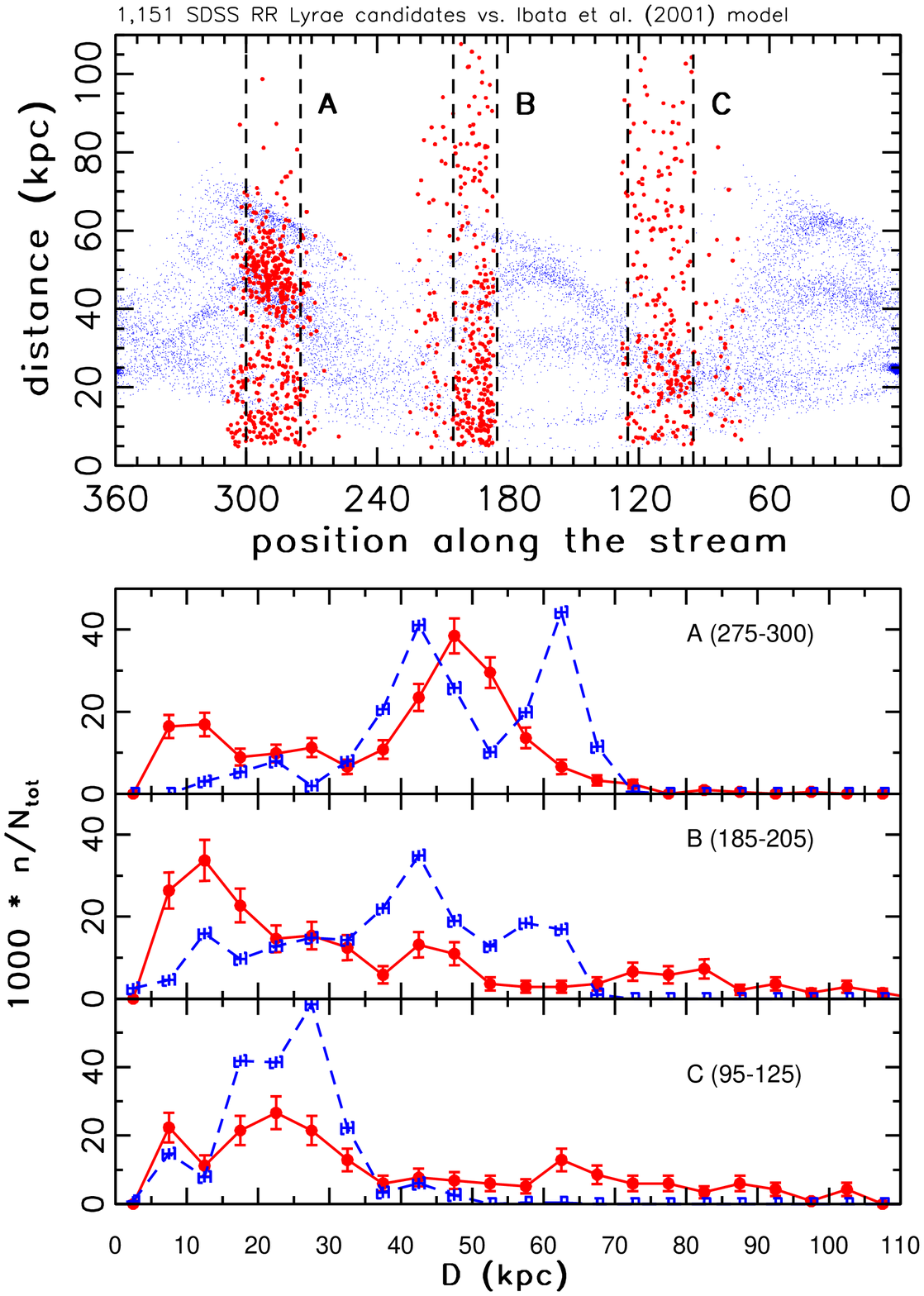}{6.5in}{0}{74}{74}{-230}{-55}
\caption{The top panel compares the data (large dots) and model (small dots,
Ibata et al. 2001) distance distributions along the great circle that 
approximates the Sgr dwarf tidal stream (see Figure~1). The bottom panel 
shows marginal distance distributions for three segments (A, B, C), 
as marked in the panels. The solid histograms correspond to data, and dashed 
histograms to models.}
\end{figure}

\begin{figure}
\plotfiddle{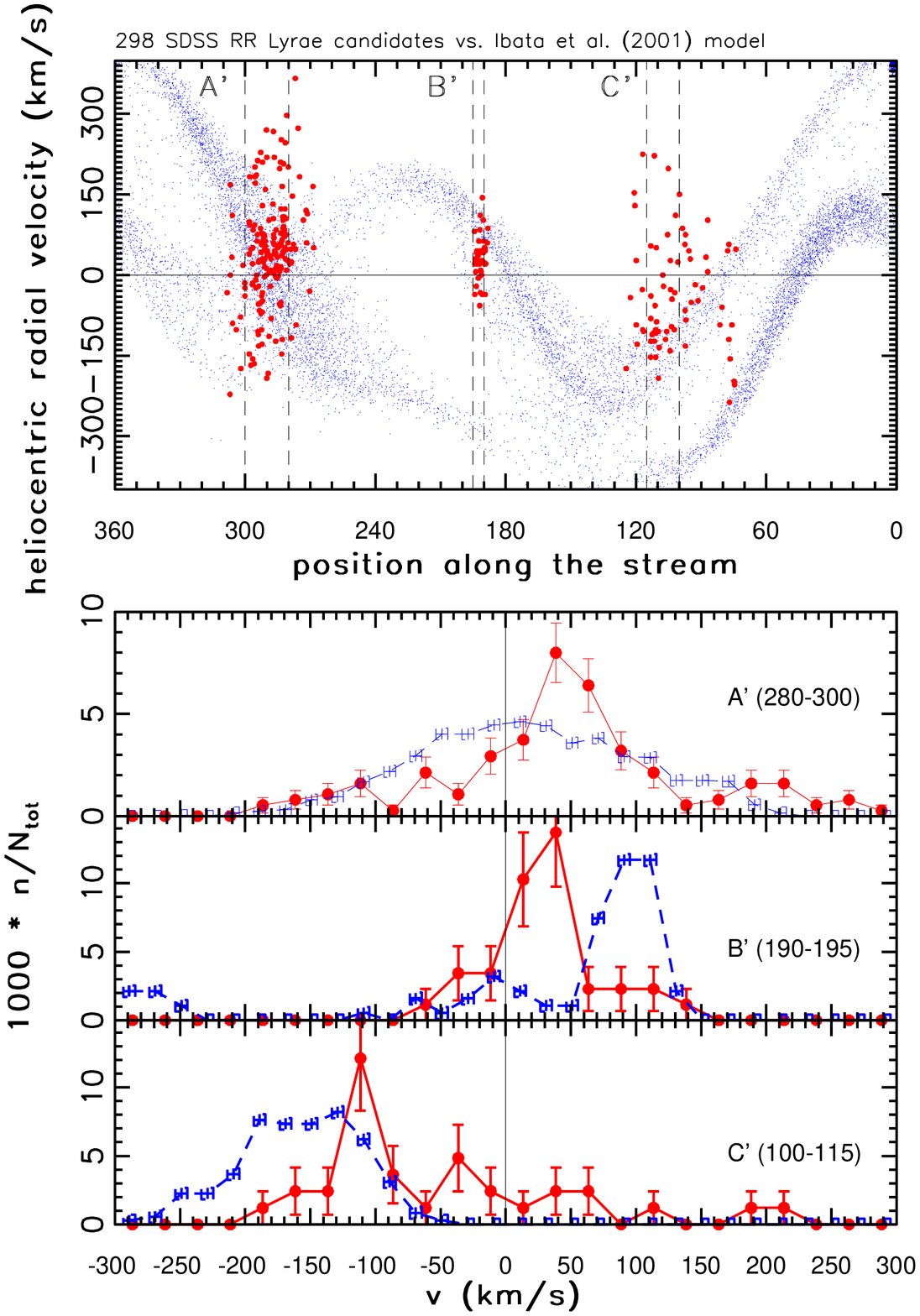}{6.5in}{0}{74}{74}{-230}{-50}
\caption{The top panel compares the data (large dots) and model (small dots,
Ibata et al. 2001) radial velocity distributions along the great circle that 
approximates the Sgr dwarf tidal stream. The bottom panel shows marginal radial 
velocity distributions for three segments (A', B', C'), as marked in the panels. 
The solid histograms correspond  to data, and dashed histograms to models.
}
\end{figure}

We select 1,157 stars that probably belong to the Sgr dwarf tidal stream
by using a 20 degree wide strip along the great circle of its orbit, as shown
in Figure~1. The distribution of distances along the stream (the origin
corresponds to the main body of Sgr dwarf galaxy) is shown as large dots 
in the top panel of Figure~3. For an analogous plot showing other data,
see Fig. 18 in Majewski et al. (2003, hereafter M03). The model prediction 
for the stellar distribution in this diagram by Ibata et al. (2001) is 
shown by small dots. The bottom panel shows marginal distance distributions 
for three segments (A, B, C), as marked in the panels. While there are 
some detailed disagreements between model and data, the model captures
well the overall appearance of the data distribution. Most notably,
there is good correspondence between multiple peaks in a given segment,
suggesting that multiple streams are projected onto the great circle
on the sky.  However, this interpretation is not conclusive because such 
clumps may also be consistent with statistical fluctuations predicted by 
Bullock, Kravtsov \& Weinberg (2001, BKW01). 
 
SDSS spectra, and thus radial velocity measurements (accurate to 
$\sim$20-30 km/s), are available for a subsample of 680 stars. The radial velocity 
distribution for the 305 stars that probably belong to the Sgr dwarf tidal stream 
is shown in Figure~4. Similar to the comparison of distance distributions, the model 
provides a good qualitative description of the data but fails to quantitatively
match the peaks in the data distributions.

\subsection{The Density Profile Outside the Sgr Dwarf Tidal Stream }

\begin{figure}
\plotfiddle{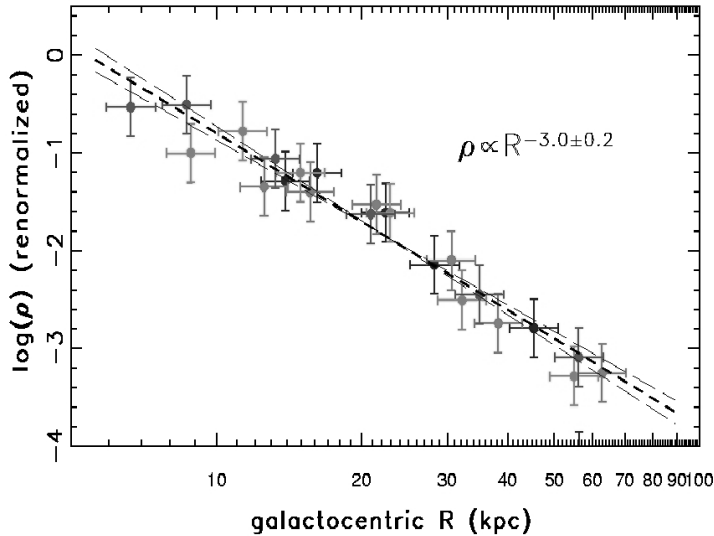}{6.5cm}{0}{50}{50}{-160}{-100}
\caption{The Galactocentric density distribution of candidate RR Lyrae
stars along four pencil beams marked in Figure~1. Outside the Sgr dwarf tidal 
stream the density profile is well described by an $R^{-3}$ power law, in agreement
with previous work (Wetterer \& McGraw 1996) which extended to $R\sim30$ kpc. 
}
\end{figure}

In order to estimate the density profile outside the Sgr dwarf tidal stream,
we select four pencil beams marked in Figure~1. They contain 400 stars 
and span a large range of orientations with respect to the Galactic center.
The Galactocentric density distribution along these four beams is shown in Figure~5. 
Remarkably, outside the Sgr dwarf tidal stream the density profile is well described 
by $R^{-3}$ power law out to 60-70 kpc from the Galactic center, in agreement 
with previous work that probed inner 30 kpc (Wetterer \& McGraw 1996, WMG96). 

\section{Discussion and Conclusions}
\vskip -0.1in
SDSS is producing a large and robust catalog of distant RR Lyrae stars which allow 
detailed investigation of the halo out to 100 kpc. Only a few years ago there
were no more than a handful of known RR Lyrae stars at distances greater than 30 kpc,
(WMG96), and now there exists a sample of 500 confirmed
halo RR Lyrae stars discovered by the QUEST survey (Vivas et al. 2001), 
and over the 3,000 probable RR Lyrae stars discussed here. These large and accurate
samples are bound to enhance our understanding of the Galactic halo.

The spatial and radial velocity distributions of candidate RR Lyrae stars discussed 
here confirm that the Sgr Dwarf tidal stream is the most pronounced feature in the 
outer halo, in agreement with M03. The new data presented here are consistent with 
multiple perigalactic passages of the Sgr Dwarf, but a more detailed analysis of models 
such as those proposed by BKW01 is needed before 
definitive conclusions can be drawn. Models for the Sgr Dwarf tidal stream provide 
a good qualitative description of the data, but there is room for quantitative improvement. 
Outside this stream, an $R^{-3}$ power law is a good description of the spatial 
distribution in the 5--60 kpc range of Galactocentric radius $R$. Further out, in 
the range 60-100 kpc, the spatial distribution appears rather clumpy (see Figure~2), 
in qualitative agreement with predictions by BKW01.
\vskip -0.1in
\phantom{x}

{\small 
Funding for the creation and distribution of the SDSS Archive has been provided by 
the Alfred P. Sloan Foundation, the Participating Institutions, the National Aeronautics 
and Space Administration, the National Science Foundation, the U.S. Department of Energy, 
the Japanese Monbukagakusho, and the Max Planck Society. The SDSS Web site is http://www.sdss.org/. 
}
\vskip -0.4in
\phantom{x}

\end{document}